\input aa.cmm
\input psfig.tex 
\voffset=1.0truecm
\baselineskip=4.1mm
\def\gr{$\gamma$-ray }
\def\grs{$\gamma$-rays }
\MAINTITLE={The galactic center arc as source of high energy $\gamma$-rays }
\AUTHOR={M. Pohl}
\INSTITUTE={Max-Planck-Institut f\"ur Extraterrestrische Physik,
Postfach 1603,
85740 Garching, Germany
}
\OFFPRINTS={\hphantom{http://www.gamma.mpe-garching} http://www.gamma.mpe-garching.mpg.de/$\sim$mkp/mkp.html}
\DATE={ }
\ABSTRACT={In this paper we discuss the radio {\it arc} at the
galactic center
to be the counterpart of the high-energy \gr source 2EG J1746-2852.

Though 2EG J1746-2852 must be regarded as true source in excess of the diffuse 
background, its position can not be determined to better than
0.3$^\circ$. The observed flux is constant within the statistical
limits and the spectrum is very hard. The lack of variability makes it
highly unlikely that any of the compact
sources in the vicinity of the Galactic Center is the counterpart of
2EG J1746-2852. This includes the peculiar source Sgr A$^\ast$ at the very
center of the Galaxy, which is often discussed to harbour a black hole
of $10^6\, M_\odot$.

Existing radio data on the {\it arc} support the view that its synchrotron
emission originates from cooling, initially monoenergetic electrons
which diffuse and convect from their sources to the outer extensions of
the {\it arc}. If the source of high-energy electrons coincides
with the sickle region (G0.18-0.04), as indicated by the radio data, then the ambient far-infrared (FIR)
photons can be up-scattered to \grs by inverse-Compton interaction 
with the young high-energy electrons. 
We solve the continuity equation for the electrons including terms
for diffusion, convection, monoenergetic injection, and the full energy loss.
With that we show that the predicted \gr emission depends mainly on
the magnetic field strength in the {\it arc} and that both the flux and
the spectrum of 2EG J1746-2852 can be well explained.
Our model shall be tested on radio data at frequencies beyond 10 GHz 
in future work.}

\KEYWORDS={Cosmic Rays - Galactic Center - Gamma Rays}

\THESAURUS={09.03.2; 10.03.1; 13.07.2; 13.07.3}

\maketitle

\titlea{The Galactic center region}

Early radio studies of the galactic center region have revealed the source 
Sgr A located at the mass and dynamical center of the Galaxy.
Sgr A has been observed at the highest angular resolution
presently obtainable for a wide range of wavelengths. The main component
of Sgr A is the extremely compact discrete source Sgr A$^{\ast}$, which is 
located close to the center of the spirallike thermal source Sgr A West.
Sgr A$^{\ast}$ is unique in the Galaxy and shows a similar behaviour
like active galactic nuclei although its luminosity is orders of magnitudes
less. VLBI observations at 43 GHz and 86 GHz indicate a size of a few
Astronomical Units or less (Krichbaum et al. 1993).
Sgr A$^{\ast}$ is variable and has an inverted synchrotron spectrum
close to $\alpha$=0.3 (S$\propto \nu^\alpha$), which may originate from
nearly monoenergetic electrons. At X-ray energies a strongly variable
source coinciding with Sgr A$^{\ast}$ has been reported 
(Skinner et al. 1987, Sunyaev et al. 1991, Churazov et al. 1994).
However, deep high-resolution observations indicate that
Sgr A$^{\ast}$ itself is a weak emitter (Predehl and Tr\"umper 1994,
for a discussion see Goldwurm et al. 1994).

Within 50 pc of Sgr A two more components are located, which are
significantly weaker and referred to as the {\it arc} and the {\it bridge}.
Recombinations lines detected from the emission of the {\it bridge}
indicate the existence of thermal material, while the detection
of polarized emission from the {\it arc} region and the absence 
of recombinations lines show the dominance of synchrotron radiation 
from this region. The magnetic field has been shown to run parallel to the
{\it arc} with a very high degree of regularity by 32 GHz observations
(Reich 1990) and 43 GHz (Tsuboi et al. 1995)
High resolution observations with the VLA have revealed
a number of long and very thin filaments superimposed on the diffuse emission
of the {\it arc} as well as isolated threads distributed throughout
the galactic center region. The isolated threads have 
normal steep radio spectra $\alpha\approx -0.5$ ($S\propto \nu^\alpha$).
The filaments in the arc have inverted spectra 
with $\alpha\simeq 0.3$ at 1 GHz (Anantharamiah et al. 1991), while they are
hardly seen at 43 GHz, implying a steep spectral index $\alpha\simeq -0.7$
between 5 GHz and 43 GHz (Sofue, Murata and Reich 1992). At 43 GHz the
filaments contribute less than 10$\%$ to the total brightness of the
{\it arc}. Given the inverted spectrum of the total emission of the
{\it arc} up to 43 GHz (Reich et al. 1988) and the high degree of
polarization at 32 GHz and 43 GHz, a substantial amount of diffuse
synchrotron emission 
with inverted spectrum is required. The {\it arc} extends northward
and southward to plumes with high degree of polarization but steeper spectrum.
These extensions may be understood as due to ageing, initially
monoenergetic electrons, which have been transported by diffusion
and convection from their sources in the {\it arc}
(Pohl, Reich and Schlickeiser 1992, abbreviated as {\bf PRS}),
although a detailed modelling suffered from systematical uncertainties in the
data set.

\titlea{EGRET results for the galactic center}

The EGRET instrument on the Compton Gamma-Ray Observatory is sensitive
to \grs in the energy range of 30 MeV to 30 GeV. The instrument design is
described in Kanbach et al. (1988) and information on
the instrument calibration may be found in Thompson et al. (1993).
For point sources a maximum
likelihood analysis (Mattox et al. 1996) is used to determine
the number of source photons, distributed according to the instrument
point spread function, in excess to a model parametrisation of the galactic
diffuse background emission.

With this technique a highly significant \gr source (2EG J1746-2852) was found
to be positionally coincident with the galactic center (Thompson et al.
1996). There is a weak preference for a source position at 
l$\approx 0.2^\circ$, especially at high \gr energies (> 1 GeV), although
the exact center position or even a
negative longitude can not be ruled out completely. 
Its integrated flux above 100 MeV is around
$10^{-6}\,$ph./cm$^2$/sec. 

Some uncertainties in the background model near the galactic 
center demand a careful analysis of the data, which in its full scope
will be published elsewhere
(Mayer-Ha\ss elwander et al., in prep.). Here we give a short overview
on the basic results which are of relevance for our modelling.

\titleb{Variability}

The \gr light curve at photon energies above 100 MeV of 2EG J1746-2852
between April 1991 and September 1993 has already been published in
Thompson et al. (1996). There is no evidence for variability during this
period of time. We have further analysed public data of Phase 3
(September 1993 to October 1994) and find no evident sign of
variability either. Therefore we conclude that the EGRET results of
Phase 1 to Phase 3 are well compatible with 2EG J1746-2852 being a constant 
emitter of \grs .

\titleb{The spectrum}
The \gr spectrum in the EGRET range has been determined on the basis
of summed data sets of Phase 1 and Phase 2 by Merck et al. (1996).
It appears to be well
represented by a hard power-law with a cut-off at a few GeV,
hence is significantly different from that of the galactic diffuse
emission, but resembles the spectra of \gr pulsars. Neglecting the 
highest energy bin (4 GeV to 10 GeV) the photon
spectral index is $s=-1.6\pm 0.09$
($I=I_0\, E^s$). 

We can also use COMPTEL data at lower energies (Strong, priv. comm.).
Since COMPTEL's limited angular resolution does not allow to discriminate
between a point source at the galactic center
and continuum emission from the galactic center molecular arm, the
final flux values were taken as upper limits for the galactic center source 
for systematic reasons. The resulting total spectrum will be shown in Fig.2.

\titleb{Discussion of EGRET results}

To summarize, 2EG J1746-2852 is a steady source of high energy \grs with 
a very hard spectrum. 

One possible counterpart is Sgr A$^{\ast}$. Standard models for the emission
of accreting matter around a massive black hole predict negligible \gr
emission (Narayan et al. 1996, Melia 1994). However, high energy emission may
arise from shock accelerated protons in the vicinity of the black hole
(Mastichiadis and Ozernoy 1994). In the latter model X-rays are produced
by synchrotron radiation of secondary electrons. These electrons can not
at the same time be responsible for the radio and near-infrared
(NIR) emission since the
cooling spectrum of secondary electrons below 100 MeV is incompatible
with the observed spectrum in the mm-range. Also, if the gas densities are
not higher than $10^7\,$cm$^{-3}$, the source should be constant in 
X-rays and \grs . 

However, Sgr A$^{\ast}$ is observed to be variable at X-ray energies and in the 
radio regime, and it is highly unlikely that it remains constant in the energy
range in which it would have to emit its peak of luminosity and in which
the physical time scales are shortest, at \gr energies. Therefore we
consider it unlikely that Sgr A$^{\ast}$ or any other compact
source in the galactic center vicinity is the counterpart to 2EG J1746-2852. 

A misinterpretation of strong diffuse emission as point source can
be excluded on the basis of the source spectrum and the incompatibility
of its structure with the spatial distribution of gas,
which forms a ridge of 1.5$^\circ$ length (Sofue 1995a,b). To account for the
observed flux the \gr emissivity per H-atom at the galactic center
molecular arm would have to be 5-10 times larger than the local value.

The \gr spectrum of 2EG J1746-2852 resembles that of a pulsar, possibly
of Geminga type. There is no radio pulsar known to be positionally
consistent with 2EG J1746-2852 (Taylor et al. 1993). Given the \gr luminosity
of Geminga a similar pulsar would have to be within a few hundred parsecs
distance to produce the observed \gr flux of 2EG J1746-2852. This
\gr source however is the only yet unidentified EGRET source which has 
a spectrum compatible with an old \gr pulsar (Merck et al. 1996), hence
it would be extremely unlikely (P$\approx 10^{-5}$) that such a hypothetic \gr 
pulsar is located positionally consistent with the Galactic center.

A remaining promising candidate for the counterpart of 2EG J1746-2852 is
the galactic
center {\it arc} in which relativistic electrons exist with unusually hard
spectrum. Based on radio data at frequencies of 10 GHz and less {\bf PRS} have
modelled the radio emission of the {\it arc} as synchrotron emission
of initially monoenergetic electrons, which propagate along the 
magnetic field and thereby loose their energy to adiabatic expansion
and radiation processes. This basic scenario is supported by recently
published 43 GHz data (Tsuboi et al. 1995), which show flat spectra
between 10 GHz and 43 GHz at -3' latitude in the {\it arc} and spectral
softening towards the outer extensions. In detail the modelling in
{\bf PRS} was hampered
by uncertainties in the contribution of localized thermal emission, especially
at the 'sickle', and in the relation of the (in single-dish data unresolved) 
filaments to the diffuse synchrotron emission, which both influence the
local relation of flux density and spectral index. Signatures of
these systematics are the weak polarization where the thermal arches cross
the {\it arc} and the shift of the peak flux density towards lower
galactic longitudes with frequency (Reich et al. 1988).
{\vskip0.4truecm
{\psfig{figure=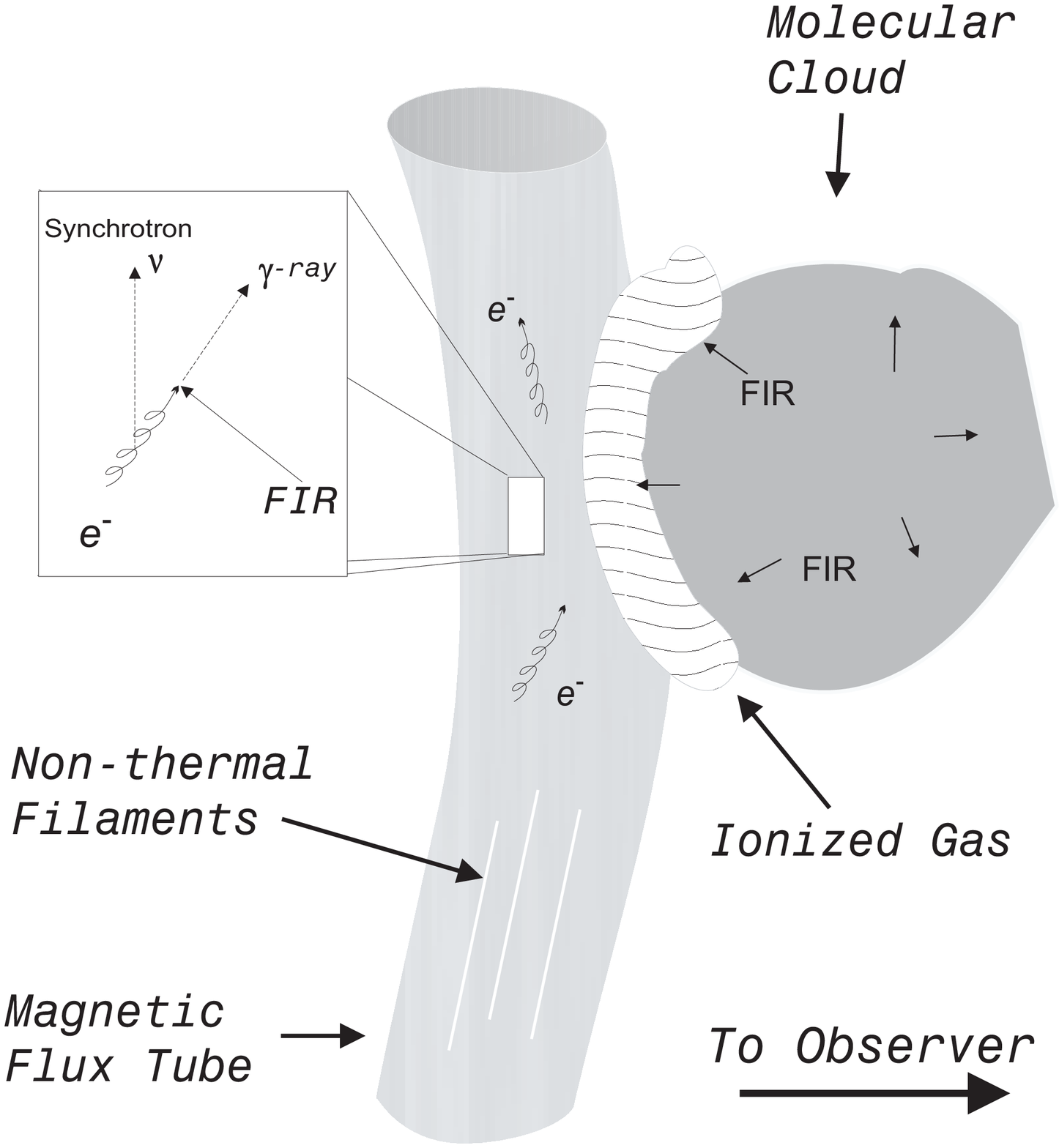,width=8.8cm}}
\figure{1}{Sketch of the basic scenario. A molecular clouds (M0.20-0.033)
presses on a magnetic flux tube (the {\it arc}) with an H$II$ region
(G0.18-0.04) as contact interface. Narrow non-thermal filaments with inverted 
radio spectrum up to 5 GHz are imbedded in the flux tube. The flux tube
itself is filled also with diffuse high-energy electrons which reveal
themselves by synchrotron radiation inverted spectrum at least up to 43 Ghz.
The most energetic electrons can up-scatter FIR photons emitted by
the molecular cloud complex to \grs .
}\vskip0.2truecm
} 

The HII region G0.18-0.04 (the 'sickle') is probably located in front of
the filaments in the {\it arc} (Lasenby et al. 1989). G018-0.04
appears to be represent the interaction zone between the dense
molecular cloud M0.20-0.033 and the magnetic field structure of
the {\it arc}. Based on the spatial coincidence of molecular clumps with
the endpoints of filaments Serabyn and Morris (1994) have argued that
these interaction regions are the acceleration sites for the 
relativistic electrons in the {\it arc}. Interestingly, the total
FIR luminosity in the G0.18-0.04/M0.20-0.033 interface region is nearly
$10^7$ L$_\odot$ (Odenwald and Fazio 1984). These FIR photons may
be inverse Compton scattered to become \grs . The basic scenario is displayed
as sketch in Fig.1, where the insert shows the radiation processes of the 
high-energy electrons: synchrotron radiation and up-scattering of the FIR
photons emitted by the molecular cloud complex M0.20-0.033.

The observed \gr spectrum 
of 2EG J1746-2852 requires an $E^{-2}$ electron spectrum which is
the natural consequence of cooling, initially monoenergetic electrons,
harmonizing with what is required to explain the synchrotron data.
In the next section we will briefly discuss the evolution of monoenergetic
particle spectra, before in section 4 we test whether the flux ratios
of \grs to synchrotron photons from the {\it arc} are compatible with the data. 

\titlea{The build-up and evolution of monoenergetic spectra}
Monoenergetic particle spectra may arise from acceleration by magnetic
reconnection (Kaastra 1985) or from the concerted action of first and second
order Fermi acceleration, synchrotron and inverse Compton losses, and escape
(Schlickeiser 1984). Here we restrict our discussion on the evolution of such 
monoenergetic spectra after the particles have escaped from the acceleration zone, independent of the acceleration process. The injection rate of 
monoenergetic particles is assumed to be constant.
Then the evolution of such particles is well described by a 
one-dimensional continuity equation (see Appendix A, eq. A1) containing
terms for spatial
diffusion, convection in an accelerated flow, and all energy loss processes.
Such an equation can be solved analytically. The solution for low-density
environments (no bremsstrahlung, no ionisation and Coulomb losses)
can be found in {\bf PRS} while the general result
is given in the Appendix A.

The volume-integrated electron spectrum is basically the inverse of the
energy loss term. The spectral index is 0 at low energies where ionisation
and Coulomb losses dominates. It changes to -1 at medium energies where
adiabatic deceleration and bremsstrahlung losses dominate while being
-2 at high energies due to the influence of synchrotron and inverse
Compton losses. The spectrum extends up to the injection energy.

The differential number density of electrons, i.e. the local spectrum,
has a more complicated behaviour. If the injection of electrons occurs only
at the symmetry point of the large-scale convection flow, then the 
spatial extent of the particle distribution is a simple
function of electron energy.
It increases as $E^{(a-1)/2}$ at high energies, 
where $a$ is the power-law exponent of the diffusion coefficient.
At medium energies the behaviour depends on the relative strength of
bremsstrahlung losses and adiabatic cooling. The spatial extent 
increases as $E^{-3}$ if only adiabatic losses are operating, is 
independent of energy if only bremsstrahlung losses are efficient, and is
in between in the general case.
At low energies the spatial extent is always independent of energy.
Therefore, the spectral index of the differential number density within the
confinement volume is 0 at low energies, -(3+a)/2 at high energies, and 
varying between 2 and -1 at medium energies.

A spatially extended source affects strongly the low energy spectra
even at larger distances from the source region, if the source
region is larger than the spatial extent of the particle distribution
at the transition energy $E_D$ between dominance of radiative losses and
dominance of
adiabatic cooling and bremsstrahlung in case of a point-like origin.
This may be part of the reason why a detailed modelling of the radio
flux density and spectral index remained inconclusive. However, this
problem does not affect the high energy \gr emission, since the target 
photons for inverse Compton scattering are not so concentrated as 
the highest energy electrons. Therefore in high energy \grs we see
all electrons (i.e. the volume-integrated spectrum), provided the 
sources of electrons are located within the region of high FIR photon density.
The particle spectral index of -2 than translates to a photon spectral index
of -1.5 for the inverse-Compton emission.

\titlea{Modelling of the $\gamma$-ray source}

From the emission measure of the thermal gas within the H$II$-region G0.18-0.04
we obtain a density
$n_e \simeq 100$ of ionized material. The molecular cloud M0.20-0.033 has 
densities $n_H \ge 10^4$
in its core as traced by CS line observations (Serabyn and Morris 1994).
The atomic and molecular gas will be at least partly external to
the {\it arc} and to the relativistic electrons confined to it.
Bremsstrahlung arising from the interaction of
relativistic electrons with this thermal matter plays only a minor role
and is not able to account for the observed \gr flux.
However, interactions with the gas have impact on the evolution of the
cooling electrons due to the corresponding energy losses. 
We have allowed for some thermal gas inside the {\it arc} with a density
small enough so that it does not dominate the electron energy loss. Our 
analytical solution of the electron transport equation requires
the energy loss terms not to vary with position. Generally a gas
density of $n\le 100\,$cm$^{-3}$ was found to be consistent with 
adiabatic losses dominating over bremsstrahlung. The examples in this section are calculated for $n= 50\,$cm$^{-3}$. Variation of this number leads to changes in
the prediction for the flux and spectral index distribution of the radio emission, 
but leaves the high energy \gr emission unaffected. The impact on the radio
emission can partly be counterbalanced by adjusting the magnetic field strength
and the electron source geometry.
A similar argument applies to the
high-energy regime where the energy losses due to inverse-Compton scattering 
should be smaller than those due to synchrotron emission which requires
$B\ge 60 \mu$G. Our example below is calculated for $B= 200 \mu$G.

The interesting \gr production process is inverse Compton scattering on the 
ambient FIR photons.
Given an FIR luminosity at the G0.18-0.04/M0.20-0.033 interface
of $10^7\ L_{\odot}$ (Morris, Davidson and Werner 1995;
Davidson, priv. comm.) the average photon energy density is around
100 eV/cm$^3$ for a spherical region of 8 pc radius.
The spectrum is assumed to be grey-body with a temperature of 50 K,
harmonizing with the surface brightness ratio at 50$\mu$m and 90 $\mu$m
(Morris, Davidson and Werner 1995).
The flux of \grs due to this process is high if the source region of electrons
coincides with or is near to the FIR emission region since the required
electrons of 100 GeV energy can not propagate far before loosing their energy.
Due to the short energy loss time scales at these energies this assumption
also implies that we see all electrons at high energies, hence the electron
spectral index is -2 up to the injection energy.

Since the photon source is external
to the {\it arc} the photon distribution cannot be regarded as isotropic.
In our modelling we approximate the molecular cloud M0.20-0.033
as point source of FIR photons located at a distance $R_0 = 10\,$pc in front
of the {\it arc}. The \gr production rate can then be calculated
in the head-on approximation for the interaction cross section.
The interested reader will find a brief description of the mathematics in
appendix B.
\vskip0.4truecm
{\psfig{figure=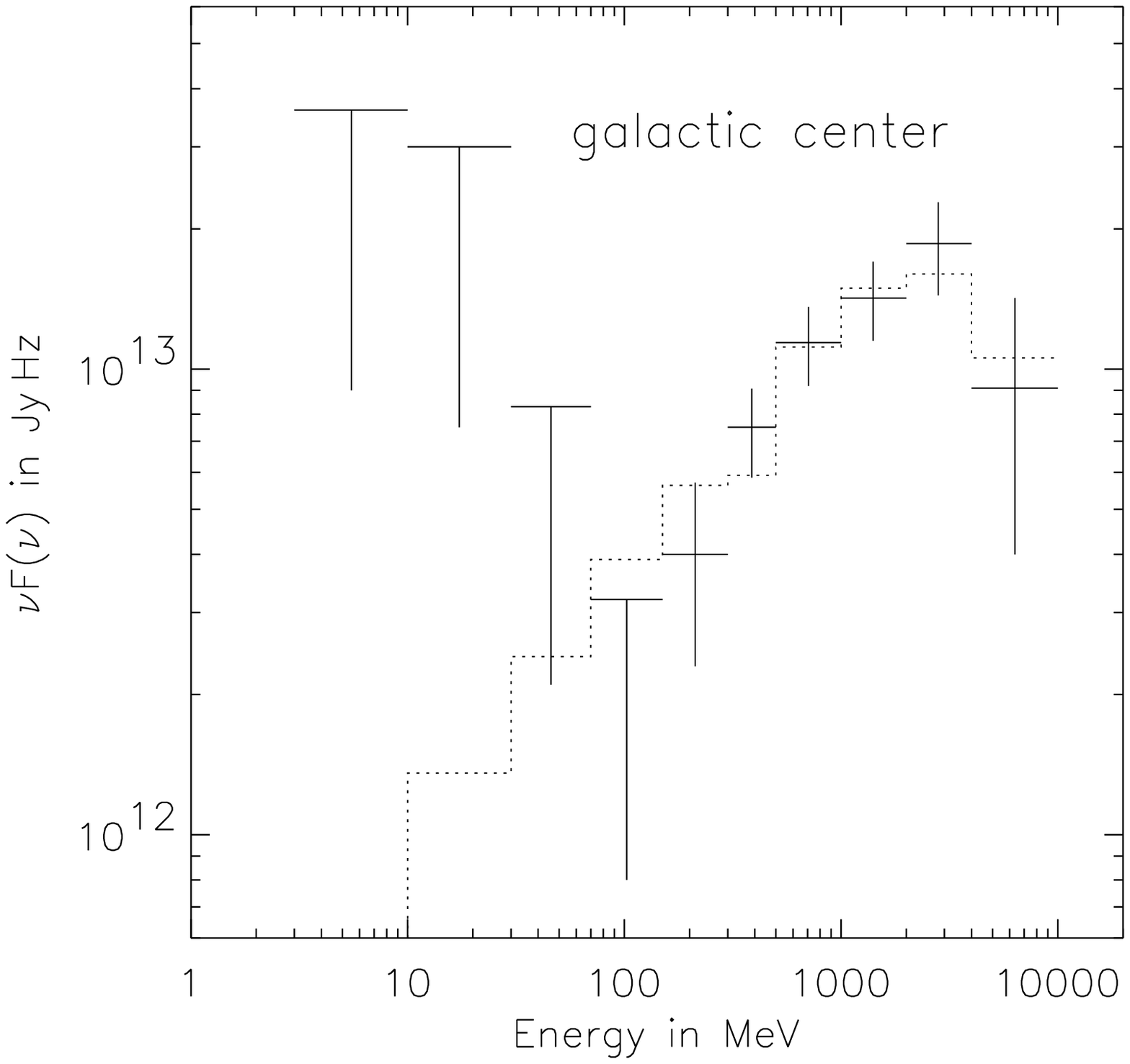,width=8.5cm,clip=}
\figure{2}{The spectrum of up-scattered FIR photons from the sickle 
region in comparison to the \gr spectrum of 2EG J1746-2852. The data points
represent the differential photon spectrum integrated over the energy bins and
multiplied with the geometric mean photon energy of the bin. The
EGRET spectrum is taken from Merck et al. (1996), where we have merged 
some of the low-energy bins. The upper limits are 1$\sigma$.
An example of the model, for which the input parameters are given in the
text, is shown as integral flux per bin in histogram mode.}
\vskip0.2truecm}

Given the synchrotron flux at the electron source region,
the magnetic field strength in the arc (not in the thin filaments!) 
and the distance $R_0$ between the {\it arc} and the FIR photon source 
the \gr flux is fixed up
to a factor of a few. The spatial distribution of synchrotron emission and 
spectral index, however, depends strongly of the extent of the
electron source region. 
So at present a detail modelling of the synchrotron emission down the
southern and northern extensions of the {\it arc} has the same problems as
before in {\bf PRS}: we cannot separate thermal and non-thermal emission,
we can hardly determine what part of the non-thermal emission at 4.75 GHz
is due to the filaments and what part is due to diffuse emission, and we cannot reliably account for the apparent widening of the {\it arc}'s cross
section with increasing latitude ({\bf PRS}). That is why we cannot
reproduce all details in the flux density distribution of the {\it arc}
(Fig.3) and in the spectral index distribution (Fig.4). 
\vskip0.4truecm
{\psfig{figure=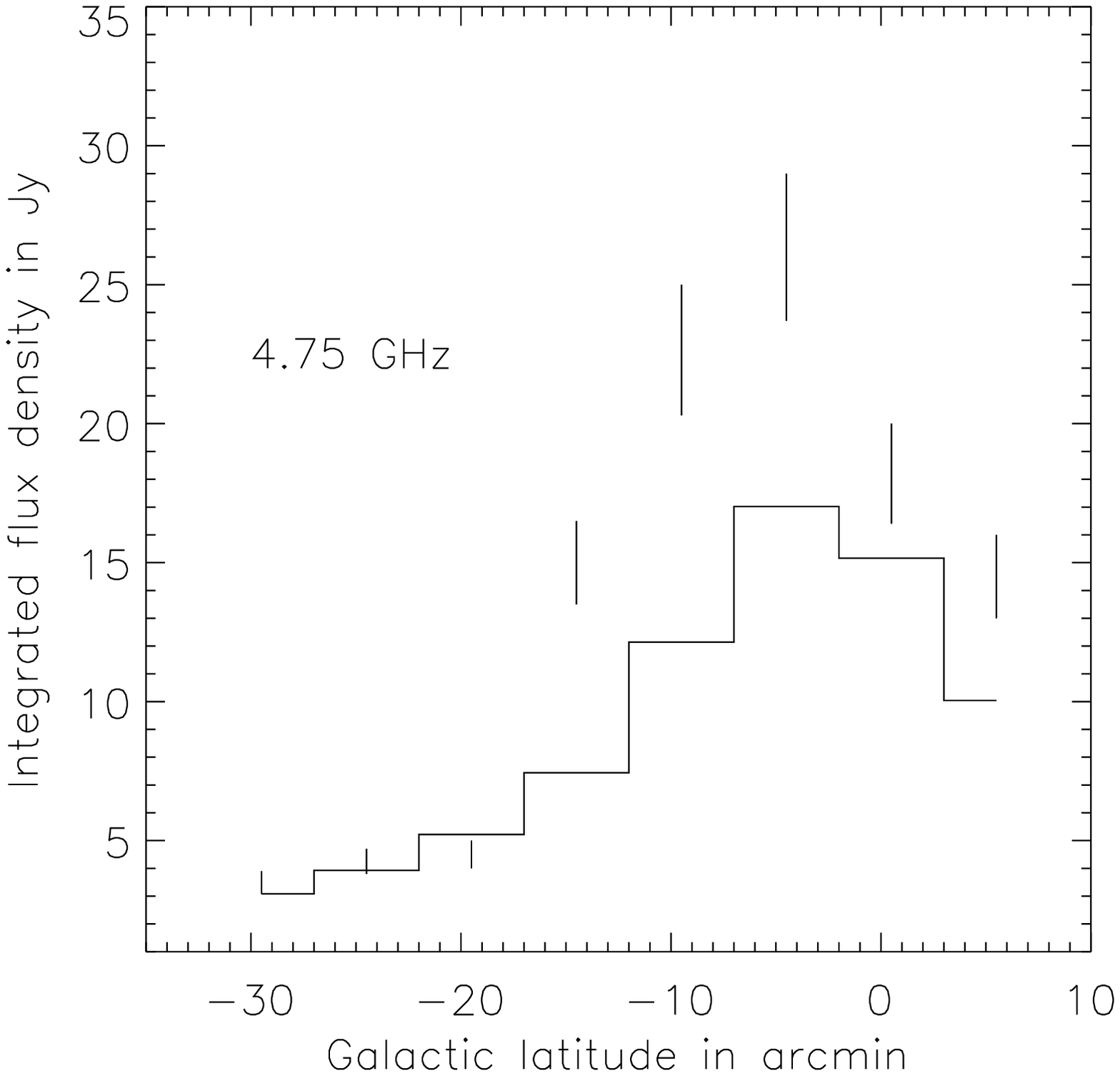,width=8.5cm,clip=}
\figure{3}{The predicted radio flux density at 4.75 GHz integrated over
the full width of the {\it arc} and bins of 5 arcmin length in comparison
to the data in {\bf PRS}, which are accordingly transformed from the
representation in their Table 1b (multiplied with a factor 2.08 times
the width of the {\it arc}).In the inner part of the {\it arc} the
model underpredicts the
flux density which is to be expected in view of contributions
from thermal emission.}
\vskip0.2truecm}
\vskip0.4truecm
{\psfig{figure=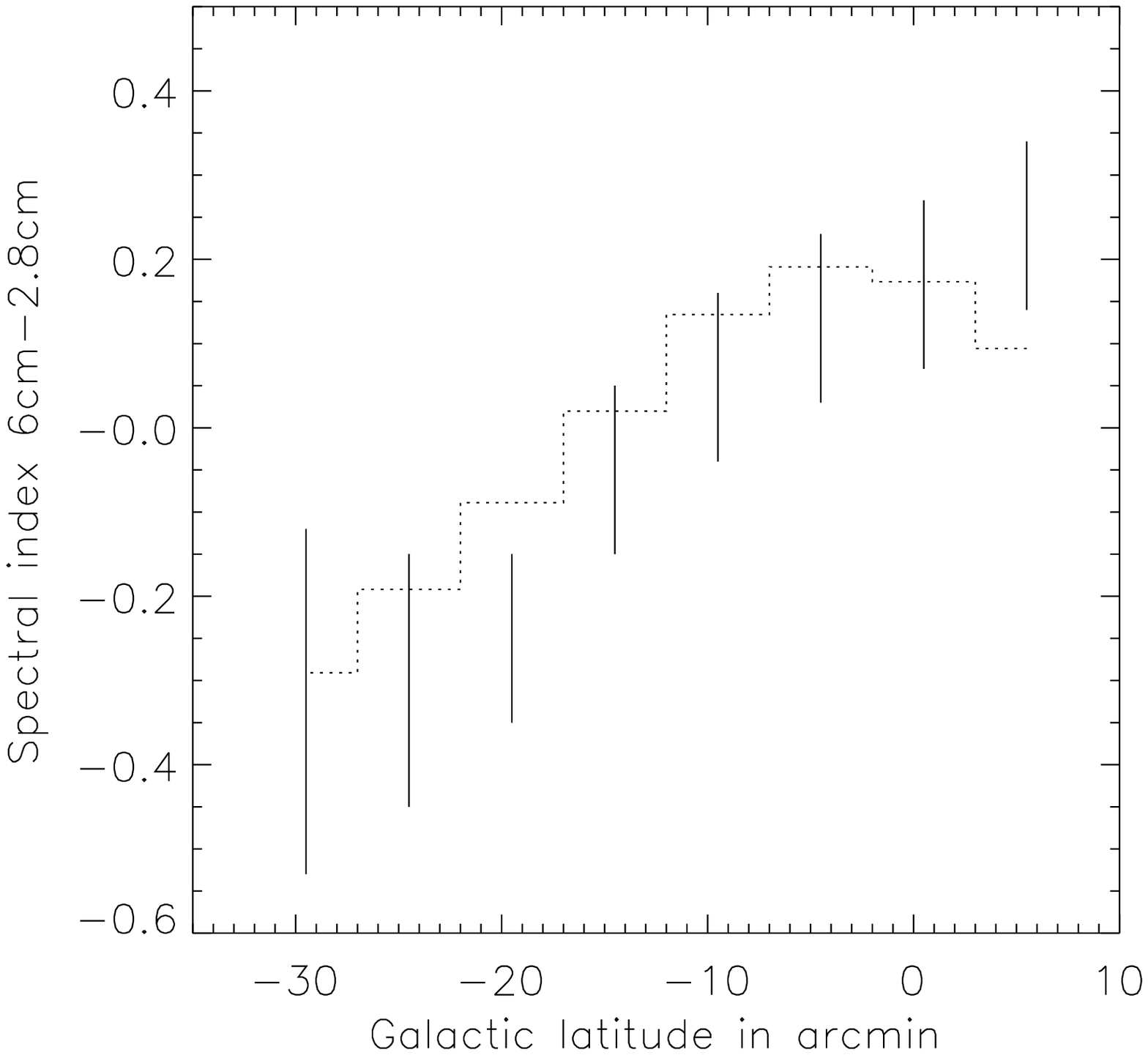,width=8.5cm,clip=}
\figure{4}{The predicted spectral index between 4.75 GHz and 10.7 GHz
averaged over the full width of the {\it arc}
and bins of 5' length in comparison to the data in {\bf PRS}.
In the inner part of the {\it arc} the observed
spectral index may be influenced by contributions
from thermal emission.}
\vskip0.2truecm}

Given the additional
contributions of thermal emission our model should underpredict the
radio flux density in the inner part of the {\it arc}.
With this kind of agreement to the radio flux density and spectral index
still the basic model reproduces both the flux and the spectrum of the
EGRET source 2EG J1746-2852 correctly, which indicates that the
{\it arc} is likely to be its counterpart. As an example we show
in Figs.2-4 the \gr spectrum, the distribution of synchrotron flux,
and the distribution of radio spectral index for a specific set of parameters.
Here the magnetic field strength is taken to be $B=200\mu G$ in
accordance to recent Zeeman splitting data (Uchida and G\"usten 1995), 
the symmetry point of the convection flow is at $b=-4'$, from which the
source region of electrons extents 6 pc to negative latitudes and 8 pc
in direction of positive latitudes. The injection energy of electrons is
$E_0=200\,$GeV, the turnover frequency ($\nu(E_D)$ in the appendix)
between dominance of adiabatic losses and radiative losses is 100 GHz, which 
then implies that adiabatic losses are nearly four times more effective
than losses due to bremsstrahlung ($\eta \simeq 4.7$ in the appendix).

Any other choice for the extent of the source region or the turnover
frequency $\nu(E_D)$ affects only the spatial distribution of
the low frequency synchrotron emission, and not the \gr emission. The 
integrated synchrotron emission scales mainly with the turn-over
frequency.

\titlea{Summary and Discussion}

In this paper we have discussed the radio {\it arc} at the galactic center
to be the counterpart of the high-energy \gr source 2EG J1746-2852.
The observed flux of this source is constant within the statistical limits.
The spectrum is very hard and resembles the spectrum of an old pulsar
(Merck et al. 1996).
Given the typical luminosity of such a pulsar its distance would have to be
much less than 1 kpc in distance to produce the observed flux. There is
no radio pulsar known compatible in position with 2EG J1746-2852.
The chance probability for a Geminga-type object coinciding spatially with
the Galactic center is around $10^{-5}$, leaving this possibility
highly unlikely, although not finally rejectable.

The lack of variability makes it highly unlikely that any of the compact
sources in the vicinity of the Galactic Center is the counterpart of
2EG J1746-2852. This includes the peculiar source Sgr A$^\ast$ at the very
center of the Galaxy, which is often discussed to harbour a black hole
of $10^6\, M_\odot$.

We have shown in section 3 and section 4 that the radio {\it arc} at the
galactic center can easily account for the observed properties
of 2EG J1746-2852 and is likely to be its counterpart.
Existing radio data on the {\it arc} support the view that its synchrotron
emission originates from cooling, initially monoenergetic electrons
which diffuse and convect from their sources to the outer extensions of
the {\it arc}. If the source of high-energy electrons coincides
with the sickle region, as indicated by the radio data, then the ambient
FIR photons can be up-scattered to \grs by inverse-Compton interaction 
with the freshly injected electrons. 

We have solved the continuity equation for the electrons including terms
for diffusion, convection, monoenergetic injection, and the full energy loss.
With that we have shown the predicted \gr emission depends mainly on
the magnetic field strength in the {\it arc} and that both the flux and
the spectrum of 2EG J1746-2852 can be well explained.
A detailed comparison of our model to the spatial variation of synchrotron
flux density and spectral index is, as in earlier work ({\bf PRS}),
still hampered by the unknown contributions of thermal emission and
the individual filaments which probably have a higher magnetic field
strength than the ambient medium. The predicted spatial distribution of 
synchrotron emission also depends strongly on the spatial extent of the
source distribution, in contrast to the prediction for the \gr emission.
Our model shall be tested on radio data at frequencies beyond 10 GHz 
in future papers.

\acknow{We like to thank Dr. R. Timmermann for his help on the FIR data
of the sickle region. We further thank Michaela Kays for preparing Fig.1.}

\appendix {A: The particle spectra including complete energy loss terms}
The continuity equation for the isotropic differential number density
N(E,z)
including terms of injection $q$, energy losses $\beta = -B(E)$,
convection with velocity $V=3 V_1 z$, and diffusion with scalar diffusion coefficient
$D=D_0 E^a$ can be solved analytically (Lerche and Schlickeiser 1981) under
some simplifying conditions.

In case of a point-like source of monoenergetic electrons with 
energy $E_0$ , i.e. $q=Q_0 \delta(E-E_0) \delta(z)$ the solution is
$$N(E,z)={{Q_0}\over {2\sqrt{\pi}}}\int_0^\infty dE'\ 
\delta (E'-E_0) 
\Theta (E'-E)$$ $$\qquad \times \  {{ {\rm exp}\left(3 V_1 \tau(E)\right)}
\over {(V_1 E + B(E))\sqrt{f}}}\ {\rm exp} \left[ {{- z^2 {\rm exp}
\left(6 V_1 \tau (E)\right)}\over {4 f}}\right]\ \eqno(A1)$$
where
$$B(E) =C \left[ 1+ {{E}\over {E_1}} + {{E^2}\over {E_2^2}}\right]$$
$$\tau(E) = \int^E du\ \left[V_1 u + B(u)\right]^{-1}$$
$$\hphantom{\tau(E)}
=\ C^{-1}\ \int^E du \ \left[1+{u\over {E_1}} + \left({u\over {E_2}}\right)^2
\right]^{-1}$$
$$f=\int_{\tau(E)}^{\tau(E')} d\tau D_0 (\tau) {\rm exp} (6V_1 \tau)\ \eqno(A2)$$
The energy loss term B(E) includes ionization and Coulomb losses, 
nonthermal bremsstrahlung, and synchrotron and inverse Compton interactions.
$\tau(E)$ can be readily integrated if $E_2 > 2\, E_1$, i.e. if $B(E)$ has only
real roots. Then
$$\tau(E) \simeq {{E_1}\over C}\ {\rm ln}\left[{{E+E_1}\over {E+E_D}}\right]\ 
,\qquad {\rm if}\quad E_D={{E_2^2}\over {E_1}} > 4\, E_1\ \eqno(A3)$$
and
$$f=\int_E^{E'} du {{D_0 u^a {\rm exp} (6V_1 \tau(u))}\over {C\left(1+
{u\over {E_1}}+{{u^2}\over {E_2^2}}\right)}}$$
The cases $E\gg E_1$ have already been solved in an earlier paper
({PRS}), however neglecting 
bremsstrahlung losses.

\vskip0.2truecm
{\bf New case:} $E\ll E_1\ll E_D$

Here 
$$\tau(E) = {{2\, E_1}\over C}\ {\rm ln}\left({{E_1}\over {E_2}}\right)\ 
\eqno(A4)$$
and 
$$f(E,E_0)\simeq {{D_0\, (7\eta -6)}\over {C\, (1-a)((a+6)\eta-6)}}\ 
E_1^{1-a}\, E_2^{2a}$$
where 
$$\eta= 1+ n_H^{-1} \left({{V_1}\over {10^{-15}\,
{\rm sec}}}\right)$$
is the ratio of adiabatic cooling to bremsstrahlung losses,
and finally
$$N(E,z)\simeq {{Q_0}\over {2 C}}\sqrt{{C(1-a)((a+6)\eta-6)}\over 
{D_0 \pi (7\eta-6)}}\left({{E_1}\over {E_2}}\right)^{6-{6\over \eta}} {{E_1^{(a-1)/2}
}\over {E_2^a}}$$
$$\qquad \times \  {\rm exp}\left[-{{z^2\, C\, 
\left({{E_1}\over {E_2}}\right)^{12-{{12}\over \eta}} 
(1-a)((a+6)\eta-6)}\over {4D_0\,
E_1^{1-a} E_2^{2a}\, (7\eta-6)}}\right]\ \eqno(A5)$$
Obviously, the spatial extent of the electron distribution does not depend
on energy, and therefore the differential number density has the same 
energy spectrum as 
the integrated particle spectrum.

The case $E_1 \ll E \ll E_D$ is given in PRS in the limit $\eta \rightarrow 
\infty$ and has to be changed accordingly.
For simplicity we set $a=0$ throughout the paper.

A simple representation of the energy-dependent scale height is
$$z_c \simeq g(E)^{-1}\sqrt{{{DE_1}\over {C \left(1+{{E}\over {E_D}}\right)}
}}\ {\rm for }\ E\ll E_0\ \eqno(A6)$$ $$
\quad {\rm where}\quad g(E)= \left[{{E+
E_1}\over {E+ E_D}}\right]^{x/2}\ ,\quad x={{6\eta-6}\over {\eta}}
\in \left[0;6\right]$$
The local particle spectrum is thus related to the integral spectrum by
$$N(E,z)\simeq {{2\,N_{tot} (E)}\over {\sqrt{\pi}\,z_c (E)}}\ {\rm exp}
\left[-{{z^2}\over {z_c^2}}\right]\ \eqno(A7)$$
where
$$N_{tot} = {{Q_0 \ \Theta (E_0 -E)}\over {C\left(1+
{E\over {E_1}} + {{E^2}\over {E_2^2}}\right)}}\ \eqno(A8)$$
In case the source region is extended to $\pm L$ around $z=0$ the 
differential number density writes
$$N(E,z,L) \simeq {{N_{tot}(E)}\over {2L}} g(E)\sum_{\pm}
{\rm erf}\left({{{{L}\over {g(E)}} \pm z}\over {z_c}}\right)\ \eqno(A9)$$
More complicated source geometries can also be accounted for
by explicit integration over the Green's function (see also Pohl and
Schlickeiser 1990).

\appendix {B: The production rate of inverse-Compton scattered photons}
Here we have to deal with the case of an isotropic electron distribution
interacting with a beamed photon distribution, i.e. all photons come from
a point-like source.

The differential production rate of \grs in the lab frame can be written as
$$S (\epsilon_s,\Omega_s, r) = c \int_0^\infty d\epsilon \oint 
d\Omega \int_1^\infty d\gamma \oint d\Omega_e (1-\beta \cos \psi ) $$ $$
\hphantom{\dot n (\epsilon_s,\Omega_s, r) =}\times\ 
n_e (\gamma,\Omega_e) n_{ph}(\epsilon,\Omega) \left( {{d\sigma}
\over {d\epsilon_s d\Omega_s}}\right)\ \eqno(B1)$$
where the energies of the incident photon ($\epsilon$) and of the
electron ($\gamma$) are in units $m_e c^2$, index $e$ refers to the electron,
and index $s$ to the final \gr. 
We use the Thomson cross section in head-on approximation
$${{d\sigma}
\over {d\epsilon_s d\Omega_s}} = \sigma_T \delta (\Omega_s - \Omega_e)
\delta \left[ \epsilon_s - \epsilon \gamma^2 (1-\beta \cos \psi )
\right] \ .\eqno(B2)$$
where the angle between incident and scattered
photon directions 
$$ \cos \psi = \mu \mu_e +\sqrt{1-\mu^2}\sqrt{1-\mu_e^2} \cos (\phi -
\phi_e)\ \eqno(B3)$$
is related to the incident photon and electron directions
($\mu_x=\cos \theta_x$).
 
While the electron distribution is isotropic
$$n_e (\gamma,\Omega_e) = {{n_e(\gamma) }\over {4 \pi}} \ \eqno(B4)$$
the target photon density is beamed
$$n_{ph}(\epsilon,\Omega) = n_{ph}(\epsilon) \delta(\phi) \delta (\mu-1)
\ .\eqno(B5)$$
After some simple algebra we obtain the \gr production rate in 
direction $\mu_s$ 
$$S (\epsilon_s,\mu_s, r) =  {{c\sigma_T}\over {4\pi}}
\int_1^\infty d\gamma \ {{n_e(\gamma) }\over {\gamma^2}}\ n_{ph}\left(
{{\epsilon_s}\over {\gamma^2 (1-\beta \mu_s)}}\right) \ \eqno(B6)$$
Let the photon source be located at a distance $R_0$ to the {\it arc} in
front of it. The total \gr emissivity 
is then derived by integration over the length of the {\it arc} where at
each position $\mu_s$ is the cosine of the angle between the
line-of-sight to the observer and the direction
to the photon source, and the target photon density is related to the
total FIR luminosity of the illuminating molecular cloud M0.20-0.033
$$ n_{ph}(\epsilon) = {{L_{FIR}\ \mu_s^2}\over {6.5\,(kT)^2\, 4\pi R_0^2\,c}}
\left({{\epsilon}\over {kT}}\right)^2 {1\over {\exp\left({{\epsilon}\over {kT}}\right) -1}}\ .\eqno(B7)$$
with the temperature $T$ fixing the grey-body spectrum.

\begref{References}

\ref Anantharamiah K.R., Pedlar A., Ekers R.D., Goss W.M.: 1991, MNRAS 249, 262

\ref Churazov E., Gilfanov M., Sunyaev R. et al.: 1994, ApJS 92, 381

\ref Goldwurm A., Cordier B., Paul J. et al.: 1994, Nat 371, 589

\ref Kaastra J.S.: 1985, PhD Thesis, University of Utrecht

\ref Kanbach G., Bertsch D.L., Favale A. et al.: 1988, Space Sci. Rev. 49, 69

\ref Krichbaum T.P., Schalinski C.J., Witzel A. et al.: 1994, in {\it
Nuclei of normal galaxies: Lessons from the Galactic Center}, Eds. R.
Genzel, Kluwer, Dordrecht

\ref Lasenby J., Lasenby A.N., Yusef-Zadeh F.: 1989, Proc. IAU Symposium 136,
Ed. M. Morris, Kluwer, Dordrecht, p.293

\ref Lerche I., Schlickeiser R.: 1981, ApJS 47, 33

\ref Melia F.: 1994, ApJ 426, 577

\ref Mastichiadis A., Ozernoy L.: 1994, ApJ 426,599

\ref Mattox J.R., Bertsch D.L., Chiang J. et al.: 1996, ApJ in press

\ref Merck M., Bertsch D.L., Dingus B.L. et al.: 1996, A\&AS submitted

\ref Morris M., Davidson J.A., Werner M.W.: 1995, in {\it Airborne Astronomy
Symposium on the Galactic Ecosystem}, Eds. M.R. Haas, J.A. Davidson and
E.F. Erickson, ASP Conference Series, Vol.73, p.477

\ref Narayan R., Yi I., Mahadevan R.: 1996, A\&AS, in press

\ref Odenwald S.F., Fazio G.G.: 1984, ApJ 283, 601

\ref Pohl M., Schlickeiser R.: 1990, A\&A 234, 147

\ref Pohl M., Reich W., Schlickeiser R.: 1993, A\&A 262, 441 ({\bf PRS})

\ref Predehl P., Tr\"umper J.: 1994, A\&A 290, L29

\ref Reich W., Sofue Y., Wielebinski R., Seiradakis J.H.: 1988, A\&A 191, 303

\ref Reich W.: Proc. IAU Symposium 140, Eds. R.Beck et al., Kluwer, Dordrecht,
p.369

\ref Schlickeiser R.: 1984, A\&A 136, 227

\ref Serabyn E., Morris M.: 1994, ApJ 424, L91

\ref Skinner G.K., Willmore A.P., Eyles C.J. et al.: 1987, Nat 330, 544

\ref Sofue Y., Murata Y., Reich W.: 1992, PASJ 44, 367

\ref Sofue Y.: 1995a, PASJ 47,527

\ref Sofue Y.: 1995b, PASJ 47, 551

\ref Strong A.W., Bennett K., Bloemen H. et al.: 1994, A\&A 292, 82

\ref Sunyaev R., Pavlinskii M., Gilfanov M. et al.: 1991, Sov. Astron. Lett. 
17(1), 42

\ref Taylor J.H., Manchester R.N., Lyne A.G.: 1993, ApJS 88, 529

\ref Thompson D.J., Bertsch D.L., Fichtel C.E. et al.: 1993, ApJS 86, 629

\ref Thompson D.J., Bertsch D.L., Dingus B.L. et al.: 1996, ApJ in press

\ref Tsuboi M., Kawabata T., Kasuga T. et al.: 1995, PASJ 47, 829

\ref Uchida K.I., G\"usten R.: 1995, A\&A 298, 473

\endref
\end
\bye